\documentclass[12pt,a4paper]{article}
\usepackage{amsmath}
\usepackage{amssymb}
\usepackage{amsthm}
\usepackage{float}
\usepackage{amsfonts}
\usepackage{graphicx}
\usepackage[]{cite}
\usepackage{float}
\usepackage{subfigure}
\usepackage{verbatim}
\usepackage[left=2cm,right=2cm,top=3cm,bottom=2.5cm]{geometry}
\usepackage[numbers]{natbib}
\usepackage[utf8]{inputenc}
\def\case#1/#2{\frac{#1}{#2}}

\def \D {\tilde{\nabla}}
\def\la {\langle}
\def\ra {\rangle}
\newcommand{\sfrac}[2]{{\textstyle{#1\over#2}}}
\def \ep {\varepsilon}
\def\tl{\tilde}
\def\rd {\displaystyle{\cdot}}
\def\ts {\textstyle}

\usepackage[usenames,dvipsnames,svgnames]{xcolor}
\usepackage[colorlinks=true,
      linkcolor=red,
      urlcolor=gray,
      citecolor=blue]{hyperref}

\def\myalign#1{%
  \def\trule{\noalign{\smallskip\hrule\medskip}}
  \def\nebc{\nearrow\bigcup}
  \def\sebc{\searrow\bigcup}
  \def\pminf{{}_{-\infty}|^{+\infty}}
  \let\Inf\infty
  \def\amp{&} 
  \vbox{\mathsurround0pt\openup1\jot
    \halign{%
      &$\displaystyle##\hfil\tabskip0pt$&\amp##\tabskip1em\crcr
      \noalign{\hrule height1pt\smallskip}#1\noalign{\smallskip\hrule height1pt}\crcr}}}

\begin{document}
\begin{center}
\textbf{On $1+3$ covariant perturbations of the quasi-Newtonian space-time in modified Gauss-Bonnet gravity}
\end{center}
\hfill\\
Albert Munyeshyaka$^{1}$,Joseph Ntahompagaze$^{2}$,Tom Mutabazi$^{1}$ and Manasse.R  Mbonye$^{2,3,4}$\\
\hfill\\ 
%$^{1}$Department of Physics, College of Science and Technology, University of Rwanda, Rwanda\;\;\; \; \hfill\\
$^{1}$Department of Physics, Mbarara University of Science and Technology, Mbarara, Uganda\;\;\; \; \;\hfill\\
$^{2}$Department of Physics, College of Science and Technology, University of Rwanda, Rwanda\;\;\; \; \;\hfill\\ 
$^{3}$ International Center for Theoretical Physics (ICTP)-East African Institute for Fundamental Research, University of Rwanda, Kigali, Rwanda
\;\;\; \; \;\hfill\\
$^{4}$ Rochester Institute of Technology, NY, USA.\\ \\\\
Correspondence:munalph@gmail.com\;\;\;\;\;\;\;\;\;\;\;\;\;\;\;\;\;\;\;\;\;\;\;\;\;\;\;\;\;\;\;\;\;\;\;\;\;\;\;\;\;\;\;\;\;\;\;\;\;\;\;\;\;\;\;\;\;\;\;\;\;\;\;\;\;\;\;\;\;\;\;\;\;\;\;
\begin{center}
\textbf{Abstract}
\end{center}
The consideration of a $1+3$ covariant approach to cold dark matter universe with no shear cosmological dust model with irrotational flows is developed in the context of $f(G)$ gravity theory in the present study. This approach reveals the existence of integrability conditions  which do not appear in non-covariant  treatments. We constructed the integrability conditions in modified Gauss-Bonnet $f(G)$ gravity basing on the constraints and propagation equations. These integrability conditions reveal the linearized silent nature of quasi-Newtonian models in $f(G)$ gravity. Finally, the linear  equations for the  overdensity and velocity perturbations of the quasi-Newtonian space-time were constructed in the context of modified $f(G)$ gravity. The application of harmonic decomposition and redshift transformation techniques to explore the behaviour of the overdensity and velocity perturbations using $f(G)$  model were made. On the other hand  we applied the quasi-static approximation to study the approximated solutions on small scales which helps  to get both analytical and numerical results of the perturbation equations. The analysis of the energy overdensity and velocity perturbations for both short and long wavelength modes in a dust-Gauss-Bonnet fluids were done and  we see that both energy overdensity  and velocity perturbations decay with redshift  for both  modes. In the limits to $\Lambda CDM$, it means $f(G)=G$ the  considered $f(G)$ model results coincide with  $\Lambda CDM$.
\\
\hfill\\
\textit{keywords:} $1+3$ covariant formalism-- cosmic acceleration-- Cosmological perturbations-- Quasi-newtonian spacetime.
\section{Introduction}\label{introduction}
Different astrophysical and cosmological observations confirm the cosmic acceleration \cite{perlmutter1998discovery,riess1998observational}. The nature of the substance behind this current cosmic accelaration is still unkown \cite{cognola2006dark,hogg2005cosmic}. It is therefore necessary to modify General Relativity(GR) for the sake of explaining this cosmic acceleration.
Currently, the most popular idea is that the energy density of the universe is dominated by dark component dubbed dark energy and to astrophysical scale, dark matter and many  proposals are drafted to explain their nature. The concordance model of cosmology appears to fit the present observations such as supernova Ia, cosmic microwave background anisotropy, large-scale structure formation, baryon accoustic oscillation and weak lensing to mention a few. This concordance model relies on the cosmological constant and cold dark matter ($\Lambda CDM$)\cite{huterer1999prospects,filippenko1998results,hinshaw2007three,seljak2005cosmological,eisenstein2005detection}. However, despite its success the $\Lambda CDM$ model is affected by its inability to address problems like the large-scale structure formation and evolution, the cosmic acceleration among many others. Addressing cosmic acceleration problem, other exotic negative pressure fluids, described  in terms of scalar fields have been taken into consideration. However at present, there is no evidence for the existence of the scalar fields responsible for the late-time accelerated expansion of the universe. This led to the searches for other viable theoretical avenues many of which are based on the idea that the dark sector originates from  modification of the gravitational interactions\cite{goheer2009coexistence}. Currently one of the most popular alternative to the concordance model is the modification of the Einstein-Hilbert action from which the dark energy may have a geometrical origin.
\\ \\
In line with that, modified gravity theories have extensively been explored to explain different cosmological scenarios including large-scale structure formation and cosmic accelaration. The most popular modified theories of gravity  include $f(R)$\cite{abebe2013large,hough2020viability,abebe2011shear,sami2018reconstructing}, $R$ being the Ricci scalar, $f(T)$ \cite{sahlu2020scalar,sami2021covariant,sahlu2021inflationary}, $T$ being the torsion scalar, $f(Q)$ \cite{sahlu2022linear,jarv2018nonmetricity,jimenez2020cosmology,flathmann2022parametrized,atayde2021can,khyllep2021cosmological}, $Q$ being nonmetricity scalar and $f(G)$\cite{li2007cosmology,cognola2006dark,rastkar2012phantom,munyeshyaka2021cosmological} models, $G$ being the Gauss-Bonnet invariant. One advantage of $f(T)$ gravity is that the field equations are second order in the metric. But the $f(T)$ gravity does not respect the local lorentz invariance unless replacing the partial derivative by lorentz covariant derivative in the definition of the torsion tensor to get new torsion scalar \cite{sahlu2020scalar}. One advantage of $f(G)$ gravity is that the Energy Momentum Tensor(EMT) decouples from the correction term. In other  modified gravity theories, the correction terms modify the EMT. The constructon of viable $f(R)$ or $f(T)$ models that are consistent with cosmological and GR constraints seems to be difficult since these modified theories of gravity give rise to a strong coupling between dark energy and a non-relativistic matter in the Einstein frame \cite{de2009construction,amendola2007f}. The goal is to explain both the early and the late-time acceleration in a geometrical way without considering the dark energy or the scalar fields\cite{capozziello2002curvature,birrell1984quantum,barth1983quantizing}. Among these different theories, a key role in this paper is played  by the Gauss-Bonnet invariant $G$. Considering a theory where both $R$ and $G$ are non-linear, exhausts the energy budget of the curvature degrees of freedom needed to modify GR. Introducing $G$ besides $R$ implies presence of two acceleration phases led by $G$ and $R$ respectively. This is possible for the non-linear combination of both $G$ and $R$  since linear $R$ reduces to GR while linear $G$ vanishes identically in four dimensional gravitational action being an invariant. The same studies  on these models have been done \cite{barth1983quantizing,de2010inevitable,capozziello2014noether,de2015cosmological,benetti2018observational} showing evidence that the Gauss-Bonnet invariant can solve some shortcomings of the original $f(R)$ and can contribute  to the explanation of the  accelerated cosmic expansion as well as the phantom behaviour, the quintessence behaviour and the transition from deceleration to acceleration phases and can also work at infrared scales \cite{capozziello2002curvature}. These combinations have some shortcomings that the ghost instabilities may arise. The work done in \cite{venikoudis2022late} presents the absence of dark energy oscillations, which indicates the advantage  of the $f(G)$ theory for the interpretation of late time phenomenology. \cite{venikoudis2022late} considered scalar field coupled with Ricci scalar and a function $f(G)$. In this study we  choose a combination of linear $R$ and non-linear $G$  to  make the $f(G)$ theories stable and able to describe the current acceleration of the universe and can be compared with the $\Lambda CDM$ for a linear $G$, that means for the case $f(G)=G$  as well as other cosmological solutions and suitable perturbation schemes\cite{nojiri2010reconstruction,cognola2007string,nojiri2006dark,de2012stability,granda2013natural}.
\\ \\
In exploration of such models, there are different approaches to study cosmological perturbations such as metric formalism \cite{song2007large,de2010cosmological,bardeen1980gauge,kodama1984cosmological,dunsby1991gauge,dunsby1992covariant} and $1+3$ covariant formalism \cite{ellis1989covariant,abebe2012covariant}. In this study, we use $1+3$ covariant formalism, a gauge-invariant formalism and the perturbation variables describe true physical degrees of freedom and no unphysical mode exists. This formalism has been extensively applied in GR \cite{ellis1989covariant}, scalar-tensor theories\cite{ntahompagaze2017f,ntahompagaze2020multifluid,ntahompagaze2018study} and different modified theories of gravity and in addition to that, one considers the spacetime with quasi-Newtonian features \cite{munyeshyaka2021cosmological,clarkson2003covariant,abebe2013large,elmardi2016chaplygin,sahlu2020scalar,sahlu2022linear}.\\ \\
Different authors have studied quasi-Newtonian cosmologies in GR limits, scalar-tensor theories and  $f(R)$ in the context of large-scale structure formation and non-linear gravitational collapse in the late universe\cite{van1997integrability,maartens1998covariant,van1998quasi,samiquasi}. The importance of investigating the quasi-Newtonian models for general relativity in cosmological context  is that there is a viewpoint in which cosmological studies can be done using Newtonian physics with the relativistic theory only needed  for examination of different observational predictions.\\ \\
Apart from that, modified theories of gravity such as $f(R)$ and $f(G)$ have been shown to exhibit more shared properties with quasi-Newtonian gravitation than GR does. A covariant approach to cold matter universe in quasi-Newtonian  cosmologies has been developed in order to derive and solve the equations governing density and velocity perturbations \cite{maartens1998covariant,sami2021perturbations,van1998quasi,maartens1998newtonian}. This approach revealed the existence of integrability conditions in GR and $f(R)$ models. In this paper we derive the integrability conditions and study the  velocity and density perturbations in the context of $f(G)$ gravity  for large-scale structure formation.\\ \\
On the other hand, the velocity and density perturbations are important to the explanation of the large scale structure formation scenario. The covariant and gauge-invariant approach to perturbations instituted  by Ellis and Bruni \cite{ellis1989covariant} will be used since it is very important that all quantities have a direct and immediate geometrical meaning  and no non-local decomposition  for example into scalar modes required and this approach provides a natural and transparent settings  to search for integrability conditions which may arise from constraint equations\cite{maartens1998covariant,maccallum1998integrability,van1997integrability,maartens1997consistency}. Furthermore,
there is  a need to study a quasi-Newtonian spacetime perturbations in $f(G)$ gravity since it will help to reveal the integrability conditions and to analyse velocity and energy density perturbation and check the large scale structure formation scenarios.
The aim of this paper is  therefore to present the framework for studying cosmological perturbations of the quasi-newtonian spacetime in $f(G)$ gravity using the $1+3$ covariant formalism.
\\ \\
In this paper, we solve the whole system of velocity and energy density perturbation equations for both GR and $f(G)$ gravity. For comparison purpose, we observe that the energy density perturbations with redshift decay for both GR and $f(G)$ gravity.  On the other side, we apply quasi-static approximation method where very slow  temporal fluctuations is considered in the perturbations for both Gauss-Bonnet energy density and momentum compared with the fluctuations of matter energy density. In this approximation, the time derivative terms of the fluctuations for Gauss-Bonnet energy density and momentum are discarded. The present work reveals that for comparison, the energy density and velocity perturbations decay with redshift  for the considered  $f(G)$ gravity model.\\ \\
The rest of this paper is organised as follows: in Sec. \ref{back} we give background field equations in modified $f(G)$ gravity and the $1+3$ Covariant description and the general propagation and constraint equations are presented. In Sec. \ref{Quasi}, we describe the Quasi-Newtonian spacetimes in modified $f(G)$ gravity and we present the integrability conditions. In Sec. \ref{pert}, we define the covariant perturbation variables for $f(G)$ gravity, derive their evolution equations and analyse their solutions.
Finally in Sec. \ref{concsec} we present the summary of the results and give conclusions.

%%%%%%%%%%%%%%%%%%%%%%%%%%%%%%%%%%%%%%%%%%%%%%%%%%%%%%%%%%%%%%%%%%%%%%%%%%%%%%%%%%%%%%%%%%%%%%%%%%%

Natural units in which $c=8\pi G_{N}=1$
will be used throughout this paper, and Greek indices run from 0 to 3.
The symbols $\nabla$, $\D$ and the overdot $^{.}$ represent the usual covariant derivative, the spatial covariant derivative, and differentiation with respect to cosmic time, respectively. We use the
$(-+++)$ spacetime signature.

\section{Background field equations and the $1+3$ covariant equations in modified $f(G)$ gravity}\label{back}

The modified Gauss-Bonnet gravity action with  matter field contribution to the Lagrangian, ${\cal L}_m\;,$ is given by\cite{li2007cosmology,uddin2009cosmological,nojiri2010reconstruction,rastkar2012phantom,munyeshyaka2022multifluid}
\begin{equation}
\mathcal {S}= \frac{1}{2} \int d^4x\sqrt{-g}\left[R+f(G)+2\mathcal {L}_m\right]\;,
\label{eq1}
\end{equation}
 with $\kappa$  is assumed to equals to $1$.  The modified Einstein equation is presented as

  \begin{eqnarray}
   && R_{\mu\nu}-\frac{1}{2}g^{\mu\nu}R= T^{m}_{\mu\nu}+\frac{1}{2}g^{\mu\nu}f-2f'RR^{\mu\nu}+4f'R^{\mu}_{\lambda}R^{\nu\lambda}-2f'R^{\mu\nu\sigma\tau}R^{\lambda\sigma\tau}_{b}\nonumber\\&&-4f'R^{\mu\lambda\sigma\nu}R_{\lambda\sigma} +2R\bigtriangledown^{\mu}\bigtriangledown_{\nu}f'-2Rg^{\mu\nu}\bigtriangledown^{2}f'-4R^{\nu\lambda}\bigtriangledown_{\lambda}\bigtriangledown^{\mu}f'\nonumber\\&&-4R^{\mu\lambda}\bigtriangledown_{\lambda}\bigtriangledown^{\nu}f'+4R^{\mu\nu}\bigtriangledown^{2}f' +4g^{\mu\nu}R^{\lambda\sigma}\bigtriangledown_{\lambda}\bigtriangledown_{\sigma}f'-4R^{\mu\lambda\nu\sigma}\bigtriangledown_{\lambda}\bigtriangledown_{\sigma}f',
   \label{eq2}
  \end{eqnarray}
where $f\equiv f(G)$ and $f'=\frac{\partial f}{\partial G}$ and $T^{m}_{\mu\nu}$ is the energy momentum tensor of the fluid matter (photons, baryons, cold dark matter, and light neutrinos) with  $G$ is the Gauss-Bonnet term given as $R^{2}-4R_{\mu \nu}R^{\mu \nu}+R_{\mu\nu\alpha \beta} R^{\mu\nu\alpha \beta}$, $R=R(g_{\mu\nu})$ is the Ricci scalar, $R^{\mu \nu}$ is the Ricci tensor and $R^{\mu\nu\alpha \beta}$ is the Riemann tensor. In this paper, we use $1+3$ covariant formalism where a spacetime is assumed to have timelike congruence thus tensors can be splitted into temporal and spatial parts with respect to the congruence \cite{gourgoulhon20073+}. The $1+3$ formalism relies on the slicing of the four dimensional spacetime by time and space(hypersurfaces) as far as GR and Einsteins equations are concerned. For congruence normal to a spacelike hypersurface, the $1+3$ formalism reduces to $3+1$ formalism \cite{park2018covariant}. The basic structure of $1+3$ formalism is a congruence of one dimensional curves mostly timelike curve, it means worldlines instead of a family of three-dimensional surfaces as in the $3+1$ formalism case \cite{gourgoulhon20073+,park2018covariant}.
 In the $1+3$ formalism, we decompose spacetime cosmological manifold $(M,g)$
 into time and space submanifold $(M,h)$ separately with a perpendicular 4-velocity field
vector $u^{a}$ so that  the 4-velocity field
vector $u^{ a}$ is defined as \cite{ellis2009republication, abebe2012covariant,ntahompagaze2018study}
\begin{equation*}
 u^{a}=\frac{dx^{a}}{d\tau},
\end{equation*}
where $\tau$ is the proper time such that $ u^{a}u_{a}=-1$. The metric $g_{ ab}$ is related to the projection tensor $h_{ ab} $ via:
 \begin{equation*}
 g_{ab} =h_{ab} - u_{a}u_{b}.
 \end{equation*}
 $u^{a}\equiv \frac{dx^{a}}{dt}$ is the $4$-velocity of fundamental observers comoving with the fluid and is used to define the
\textit{covariant time derivative} for any tensor
${S}^{\alpha..\beta}{}_{\gamma..\delta} $ along an observer's worldlines:
\begin{equation}
 \dot{S}^{\alpha..\beta}{}_{\gamma..\delta}{} = u^{\lambda} \bigtriangledown_{\lambda} {S}^{\alpha..\beta}{}_{\gamma..\delta}\;.
\end{equation}

 In the $4$-velocity field $u_{ a}$ , the Ehlers-Ellis approach gives fully
covariant quantities and equations with transparent physical and geometric meaning \cite{van1998quasi,ellis2009republication,abebe2017integrability,ntahompagaze2018study}.
The energy-momentum tensor of matter fluid forms is given by \cite{ellis2009republication, abebe2012covariant,ntahompagaze2018study}
\begin{equation}
T_{\mu\nu} = \rho u_{\mu}u_{\nu} + ph_{\mu\nu}+ q_{\mu}u_{\nu}+ q_{\nu}u_{\mu}+\pi_{\mu\nu}\;,
\label{eq4}
\end{equation}
where $\rho$, $p$, $q_{\mu}$ and $\pi_{\mu\nu}$ are the energy density, isotropic
pressure, heat flux and anisotropic pressure of the fluid respectively.
The first covariant derivative of $u^a$ can be split into its
irreducible parts as
\begin{equation}
\bigtriangledown_{\alpha} u_{\beta}=-A_{\alpha} u_{\beta}+\frac{1}{3}\Theta h_{\alpha\beta}+\sigma_{\alpha\beta}+\epsilon_{\alpha \beta \gamma}\omega^{\gamma},
\end{equation}
where $A_{\alpha}\equiv \dot{u}_{\alpha}$, $\Theta\equiv \tilde{\bigtriangledown}_{\alpha} u^{\alpha}$,
$\sigma_{\alpha \beta}\equiv \tilde{\bigtriangledown}_{\langle \alpha}u_{\beta \rangle}$ and $\omega^{\alpha}\equiv\epsilon^{\alpha \beta \gamma}\tilde{\bigtriangledown}_{\beta} u_{\gamma}$.
The quantities $\pi_{\mu\nu}$, $q^{\mu}$, $\rho$ and $p$ are reffered  to as dynamical quantities which can be obtained from the energy momentum tensor $T_{\mu\nu}$ as
\begin{eqnarray}
 &&\rho=T_{\mu\nu}u^{\mu}u_{\nu},\\
&& p=-\frac{1}{3}h^{\mu\nu}T_{\mu\nu},\\
&& q_{\mu}=h^{\sigma}_{\mu}u^{\rho}T_{\rho\sigma},\\
&& \pi_{\mu\nu}=h^{\rho}_{\mu}h^{\sigma}_{\nu}+ph_{\mu\nu}.
\end{eqnarray}
The quantities $\sigma_{\mu\nu}$, $\varpi_{\mu\nu}$, $\theta$ and $A_{\mu}$ are reffered as kinematical quantities.
The consideration of  standard matter fields (dust, radiation, etc) and Gauss-Bonnet contributions leads  us to define  the total energy density, isotropic  pressures, heat flux and the projected symmetric trace-free anisotropic stress as
\begin{equation}
 \rho=\rho_{m}+\rho_{G},~ p=p_{m}+p_{G},
 ~q^{a}=q_{m}^{a}+q_{G}^{a},~ \pi^{ab}=\pi_{m}^{ab}+\pi_{G}^{ab},
 \label{eq17}
\end{equation}
where \cite{li2007cosmology,garcia2011energy,munyeshyaka2021cosmological,venikoudis2022late}
\begin{eqnarray}
 && \rho_{G}=\frac{1}{2}(f'G-f)-24f''\dot{G},\\
 && p_{G}=\frac{1}{2}(f-f'G)+\frac{G\dot{G}}{3H}f''+4H^{2}\ddot{G}f''+4H^{2}\dot{G}^{2}f''',\\
 && q_{G}^{a}=-\frac{4}{9}\theta^{2}\left(f'''\dot{G}\bigtriangledown_{a}G+f''\bigtriangledown_{a}\dot{G}-\frac{\theta}{3}f''\bigtriangledown_{a}G\right),\\ \label{eq20}
&& \pi_{G}^{ab}=\frac{G}{2\theta}f''\bigtriangledown_{a}\bigtriangledown_{b}G+2\left(f''\ddot{G}+f'''\dot{G}^{2}+\frac{1}{3}\theta f''\dot{G}\right)\pi_{ab}+\frac{G}{2\theta}f''\dot{G}  \sigma_{ab}\nonumber\\
&&\quad\quad+\left(\frac{4}{3}\theta \dot{G}f''-4\ddot{G}f''-4f'''\dot{G}^{2}\right)E_{ab}.
 \label{eq21}
\end{eqnarray}
For a spatially flat FRW universe we have
\begin{equation}
 ds^{2}=-dt^{2}+a^{2}\left(dx^{2}+dy^{2}+dz^{2}\right),
\end{equation}
  and  the equation corresponding to the Friedmann equation is presented as follows:
\begin{eqnarray}
 &&3H^{2}=\frac{1}{2}\left( Gf'-f-24\dot{G}H^{3}f''\right)+\rho_{m}\;,
 \label{eq23} \\
 &&  G=24H^{2}(\dot{H}+H^{2})\;,
 \label{eq24}\\
  &&  R=6(\dot{H}+2H^{2})\;,
  \label{eq25}
\end{eqnarray}
where $H=\frac{\dot{a}}{a}$ is the Hubble parameter and $a$ is the scale factor.
The  energy density and pressure in the modified Gauss-Bonnet gravity are presented as
\begin{eqnarray}
&& \rho=3H^{2}\;,
 \label{eq11}\\
&& p=-(3H^{2}+2\dot{H})\;.
 \label{eq12}
\end{eqnarray}
The  matter is considered as barotropic perfect fluid such that $p=w\rho$, where $w$ is the equation of state parameter.
    \subsection{Propagation equations}
The covariant linearised evolution equations are given by \cite{bardeen1980gauge,van1998quasi,abebe2017integrability,samiquasi,abebe2015irrotational,ntahompagaze2020multifluid}
\begin{eqnarray}
&& \dot{\theta}+\frac{1}{×3}\theta^{2}+\frac{1}{2}(\rho+p)-\tilde{\bigtriangledown}^{a}A_{\mu}=0,
\label{eq38}\\
&& \dot{\sigma}_{\mu\nu}+\frac{2}{3}\theta\sigma_{\mu\nu}+E_{\mu\nu}+\frac{1}{2}\pi_{\mu\nu}-\tilde{\bigtriangledown}_{a}A_{\nu}=0,
 \label{eq39}\\
&& \dot{\varpi}^{\mu}+\frac{2}{3}\theta\varpi^{\mu}-\frac{1}{2}\eta^{\mu\nu\sigma}\tilde{\bigtriangledown}_{a}A_{\nu}=0,
 \label{eq40}\\
&& \frac{1}{2}\dot{\pi}_{\mu\nu}+\frac{1}{6}\theta \pi_{\mu\nu}-\frac{1}{2}(\rho+p)\sigma_{\mu\nu}-\frac{1}{2}\tilde{\bigtriangledown}_{a}q_{\mu}-\dot{E}_{\mu\nu}-\theta E_{\mu\nu}+\eta^{\rho\sigma}\tilde{\bigtriangledown}_{a}H^{\nu}_{\sigma}=0,
\label{eq41}\\
&& \dot{H}_{\mu\nu}+\theta H_{\mu\nu}+\eta^{\rho\sigma}\tilde{\bigtriangledown}_{\rho}E^{\nu}_{\sigma}+\frac{1}{2}\eta^{\rho\sigma}\tilde{\bigtriangledown}_{\rho}\pi^{\nu}_{\sigma}=0.
\end{eqnarray}
These equations will be used in the derivation of integrability conditions and the perturbation equations.
\subsection{Constraints Equations}
The covariant linearised constraint equations  are given by \cite{bardeen1980gauge,van1998quasi,abebe2017integrability,samiquasi,ntahompagaze2020multifluid}
\begin{eqnarray}
 &&\eta^{\rho\sigma}\tilde{\bigtriangledown}_{\rho}\varpi^{\nu}_{\sigma}=0,\\
&& q^{\mu}=-\frac{2}{3}\tilde{\bigtriangledown}_{\mu}\theta+\tilde{\bigtriangledown}^{\nu}\sigma_{\mu\nu}+\tilde{\bigtriangledown}^{\nu}\varpi_{\mu\nu}\label{eq44},\\
&& H_{\mu\nu}=\eta^{\rho\sigma}[\tilde{^{\mu}\bigtriangledown}_{\rho}\sigma^{\rho}_{\nu}+\tilde{\bigtriangledown} ^{\mu}\varpi_{\nu}],
\label{eq45}\\
&& \tilde{\bigtriangledown}^{\nu}E_{\mu\nu}=\frac{1}{2}[\tilde{\bigtriangledown}^{\mu}\pi_{\mu\nu}+\frac{2}{3}\theta q_{\mu}+\frac{2}{3}\tilde{\bigtriangledown}_{\mu}\rho],
\label{eq46}\\
&& \tilde{\bigtriangledown}^{\nu}H_{\mu\nu}=\frac{1}{2}\left[\tilde{\bigtriangledown}_{\sigma}q_{\rho}+(\rho+p)\varpi_{\sigma\rho}\right]\eta^{\mu\nu\sigma}.
\label{eq47}
\end{eqnarray}
where $\eta_{\mu \nu \sigma \rho}$ is the covariant permutation tensor, and $E_{\mu \nu}$ and $H_{\mu\nu}$ are respectively the electric and magnetic parts of the Weyl tensor responsible for tidal forces and gravitational waves.
The projected Ricci scalar $\tilde{R}$ into the hypersurfaces orthogonal to $u^{a}$ is given as
\begin{equation}
 \tilde{R}=2\rho-\frac{2}{3}\theta^{2}.
 \label{eq48}
\end{equation}
where $\theta= 3H$.
The conservation equation is
\begin{equation}
 \dot{\rho}+(\rho+p)\theta+\tilde{\bigtriangledown}^{a}q_{a}=0.
 \label{eq49}
\end{equation}
The other important equation for $\dot{u}$ is given as\cite{ellis2009republication, abebe2012covariant,ntahompagaze2018study}
\begin{equation}
 \dot{q}_{a}+\frac{4}{3}\theta q_{a}+(\rho+p)\dot{u}_{a}-\tilde{\bigtriangledown}_{a}p+\tilde{\bigtriangledown}^{b}\pi_{ab}=0.
\end{equation}
In the  FRW background, the dynamic, kinematic and gravito-electromagnetic equations become
\begin{equation}
 \tilde{\bigtriangledown}_{a}\rho=0=\tilde{\bigtriangledown}_{a}p=\tilde{\bigtriangledown}_{a}\theta, A_{a}=0=\varpi_{a}=q_{a}, \pi_{ab}=0=\sigma_{ab}=E_{ab}=H_{ab}.
\end{equation}
As we will assume, the spatially flat universe, no curvature on large scale, therefore $\tilde{R}=0$, hence eq. \ref{eq48} deduces to
\begin{equation}
 \frac{1}{3}\theta^{2}=\rho,
\end{equation}
which is the Friedman equation in standard general relativity. The Raychaudhuri equation in modified Gauss-Bonnet gravity is presented as
 \begin{equation}
  \dot{\theta}+\frac{1}{3}\theta^{2}+\frac{1}{2}\left[(1+3w_{m})\rho_{m}+f-f'G+\left(\frac{G\dot{G}}{H}-24\dot{G}+12H^{2}\ddot{G}\right)f''+12H^{2}\dot{G}^{2}f'''\right]=0.
 \end{equation}
\section{Quasi-Newtonian spacetimes in modified $f(G)$ gravity}\label{Quasi}
\subsection{Introduction}
The comoving $4$-velocity $\tilde{u}^{a}$ can be defined  in the linearised form as,
\begin{equation}
 \tilde{u}^{a}=u^{a}+v^{a},
\end{equation}
$ u_{a}u^{a}=0$, $v_{a}v^{a}\ll 1,$
 so there is a change in the dynamical, kinematical and gravito-electromagnetic quantities.
Here $v^{a}$ is the relative velocity of the comoving frame with respect to the observer in the quasi-Newtonian frame, defined such that it vanishes in the background. Quasi-newtonian spacetime is known to be irrotational, shear free dust characterised by \cite{abebe2017integrability,maartens1998covariant}
\begin{equation}
 p_{m}=0, q_{a}^{m}=\rho_{m}v_{a}, \pi_{ab}^{m}=0, \varpi_{a}=0, \sigma_{ab}=0.
 \label{eq56}
\end{equation}
The shear free $\sigma_{ab}=0$, irrotational $\varpi_{ab}=0$ and the gravito-electromagnetic constraint equation (eq. \ref{eq45}) result in the silent constraint
\begin{equation}
 H_{ab}=0,
\end{equation}
thus there is no gravitational radiation, implying the term 'Quasi-Newtonian'.
Due to  vanishing of the shear,  eq. \ref{eq41} becomes
\begin{equation}
 E_{ab}=\tilde{\bigtriangledown}_{a}A_{b}-\frac{1}{2}\pi_{ab},
 \label{eq58}
\end{equation}
and using the identity in Eq. (\ref{a0}) for any scalar $\varphi$, Eq. \ref{eq40} can be simplified as
\begin{equation}
 \tilde{\bigtriangledown}_{a}A_{b}=0,
 \label{eq59}
\end{equation}
 therefore \cite{van1998quasi,abebe2017integrability}
\begin{equation}
  A_{a}=\tilde{\bigtriangledown}_{a}\varphi,
  \label{eq60}
\end{equation}
where $\varphi$ is the covariant relativistic generalisation of the Newtonian potential.
Eqs \ref{eq47} and \ref{eq56}  show that $q_{a}$ is irrotational and thus so $v_{a}$
\begin{equation}
 \tilde{\bigtriangledown}_{b} ~q_{a}=0.
\end{equation}
It follows that for a vanishing vorticity, there exists a velocity potential $\psi$ such that
\begin{equation}
 v_{a}=\tilde{\bigtriangledown}_{a}\psi.
\end{equation}
 In general, if the silent constraint $H_{ab}=0$ is imposed, the linear models are consistent but the non-linear models are not consistent \cite{van1998quasi,abebe2017integrability}. Thus the usual approach to the integrability conditions for quasi-newtonian cosmologies follows from showing that these models are in fact a subclass of the linearised silent models. This is possible when one considers the transformation between the quasi-Newtonian frame and comoving frame. Therefore, for $u_{a}$ and $\tilde{u}_{a}$ in non-relativistic relative motions, the transformed linealised
 kinematic, dynamic and gravito-electromagnetic quantities are presented as  \cite{bardeen1980gauge,van1998quasi,abebe2017integrability,samiquasi,abebe2015irrotational,van1997integrability}
\begin{eqnarray}
 &&\tilde{\theta}=\theta+\tilde{\bigtriangledown}^{a}v_{a},\\
 && \tilde{A}_{a}=A_{a}+\dot{v}_{a}+\frac{1}{3}\theta v_{a},\\
&& \tilde{\varpi}_{a}=\varpi-\frac{1}{2}\eta_{abc}\tilde{\bigtriangledown}^{b}v^{c},\\
 &&\tilde{\sigma}_{ab}=\sigma_{ab}+\tilde{\bigtriangledown}_{a}v_{b},\\
 && \tilde{\rho}=\rho, \tilde{p}=p, \tilde{\pi}_{ab}=\pi_{ab}, \tilde{q}_{a}=q_{a}-(\rho+p)v_{a},\\
 &&\tilde{E}_{ab}=E_{ab}, \tilde{H}_{ab}=H_{ab}.
\end{eqnarray}
respectively. Here $u_{a}$ is the quasi-newtonian frame and $\tilde{u}_{a}$ the comoving frame. It follows that
\begin{eqnarray}
 && \tilde{p}=0, \tilde{q}^{a}=0, \tilde{\pi}_{ab}=0,\\
&& \tilde{\varpi}=0, \tilde{\sigma}_{ab}=\tilde{\bigtriangledown}_{a}v_{b},\\
&& \tilde{H}_{ab}=0, \tilde{E}_{ab}=E_{ab}.
\end{eqnarray}
Eq. \ref{eq21} reduces to
\begin{equation}
\pi_{G}^{ab}=\frac{G}{6H}f''\tilde{\bigtriangledown}_{a}\tilde{\bigtriangledown}_{b}G.
 \label{eq72}
\end{equation}
\subsection{First Integrability condition in modified $f(G)$ gravity\\}
The time derivative of equation eq. \ref{eq58}  and using eq. \ref{eq60}, eq. \ref{eq72} and eq. \ref{a0} yields
\begin{equation}
 \tilde{\bigtriangledown}_{a}\tilde{\bigtriangledown}_{b}(\frac{1}{3}\theta+\dot{\psi}-\frac{G\dot{G}f''}{6H})+(\frac{1}{3}\theta+\dot{\psi}-\frac{G\dot{G}f''}{6H})\tilde{\bigtriangledown}_{a}\tilde{\bigtriangledown}_{b}\psi=0.
 \label{eq73}
\end{equation}
This is the first integrability condition for quasi- Newtonian cosmologies in $f (G )$ gravity
and it is a generalisation of the one obtained in \cite{maartens1998covariant,sami2021covariant}. Eq. \ref{eq73} reduces into an identity for the
generalized Van-Ellis condition.
Therefore one has  \cite{abebe2017integrability, sami2021covariant,maartens1998covariant}
\begin{equation}
 \frac{1}{3}\theta+\dot{\psi}=\frac{G\dot{G}f''}{6H}.
 \label{eq74}
\end{equation}
\subsubsection{Modified Poisson equation\\}
By making  time derivative of Eq. \ref{eq74} and using eq. \ref{eq38} and eq. \ref{eq49}, the covariant modified Poisson equation in $f(G)$ gravity can be derived and presented as follows

\begin{eqnarray}
 &&\bigtriangledown^{2}_{a}\psi =-(3\ddot{\psi}+\theta \dot{\psi})+\frac{1}{2}\rho_{m}\left(1+\frac{G\dot{G}f''}{6H^{2}}\right)+\frac{1}{2}(f-Gf')\left(1+\frac{G\dot{G}f''}{6H^{2}}\right)\nonumber \\
 && \quad\quad+\frac{G\dot{G}f''}{2H}\left(2+\frac{2\theta}{3}-\frac{3G\dot{G}f''}{H^{2}}-2H^{2}\dot{G}f''+2H\ddot{G}f''+2H\dot{G}^{2}f'''\right)\nonumber\\
 && \quad\quad+\frac{\dot{G}^{2}f''}{2H}+\frac{G\dot{G}^{2}f'''}{2H}-6H^{2}\left(\ddot{G}f''-H\dot{G}f''+\dot{G}^{2}f'''\right).
 \label{eq75}
\end{eqnarray}
In the limit to the $\Lambda$CDM, it means $f(G)=G$,  from Eq. \ref{eq75} we obtain the  covariant Poisson equation \cite{abebe2017integrability, sami2021covariant,maartens1998covariant}
\begin{equation}
 \bigtriangledown^{2}_{a}\psi -\frac{1}{2}\rho_{m}+(3\ddot{\psi}+\theta \dot{\psi})=0,
 \label{eq76}
\end{equation} which coincides with $\Lambda$CDM.
\subsubsection{The evolution equation for the $4$-acceleration $A_{a}$\\}
Taking the spatial gradient of Eq. \ref{eq74}, we have
\begin{equation}
 \tilde{\bigtriangledown}_{a}\dot{\psi}=-\frac{1}{3}\tilde{\bigtriangledown}_{a}\theta+\frac{Gf''}{6H}\tilde{\bigtriangledown}_{a}\dot{G}+\dot{G}\tilde{\bigtriangledown}_{a}(\frac{Gf''}{6H}).
\end{equation}
 Using the identity \ref{a4}, eq. \ref{eq60} together with the shear free constranint eq. \ref{eq44} the evolution equation for $4$-acceleration can be found and presented as  \begin{eqnarray}
 &&\dot{A}_{a} +(\frac{2}{3}\theta-\frac{G\dot{G}}{6H}f'')A_{a}+(\frac{1}{2}+\frac{3\dot{G}f''}{8\theta^{2}})\rho_{m}v_{a}-\left(\frac{G\dot{G}(f'')^{2}}{3}+\frac{2}{9}\theta^{2}f''+\frac{Gf''}{2\theta}\right)\tilde{\bigtriangledown}_{a}\dot{G}\nonumber\\
 && \quad\quad-\left[\left(f'''\dot{G}-\frac{\theta}{3}f''\right)\left(\frac{2}{9}\theta^{2}+\frac{G\dot{G}f''}{3} \right)+\frac{\dot{G}}{2\theta}\left(f''+Gf'''\right)\right]\tilde{\bigtriangledown}_{a}G=0.
 \label{eq78}
\end{eqnarray}
In the limit to the $\Lambda$CDM, it means $f(G)=G$,  from Eq. \ref{eq79} we obtain \cite{maartens1998covariant}
\begin{equation}
 \dot{A}_{a} +\frac{2}{3}\theta A_{a}+\frac{1}{2}\rho_{m}v_{a}=0,
 \label{eq79}
\end{equation} which coincides with $\Lambda$CDM.
To check for the consistency of the constraint Eq. \ref{eq58} on any spatial hypersurface of constant time $t$, we take the divergence of Eq. \ref{eq58} which will help us to get the second integrability condition.
\subsection{ Second integrability condition in modified $f(G)$ gravity\\}
Taking the gradient of eq. \ref{eq58}, we have
\begin{equation}
 \tilde{\bigtriangledown}_{a}E_{ab}=-\frac{1}{2}\tilde{\bigtriangledown}_{b}\pi_{ab}+\frac{1}{2}\tilde{\bigtriangledown}^{2}(\tilde{\bigtriangledown}_{a}\psi)+\frac{1}{6}\tilde{\bigtriangledown}_{a}\tilde{\bigtriangledown}^{c}(\tilde{\bigtriangledown}_{a}\psi)+\frac{1}{3}(\rho-\frac{1}{3}\theta^{2})\tilde{\bigtriangledown}_{a}\psi.
\end{equation}
Using constriant equation eq. \ref{eq46}, eq. \ref{eq17}  together with the divergence of eq. \ref{eq21}, identity \ref{a21} and eq. \ref{eq44}, we therefore have
\begin{eqnarray}
 &&\tilde{\bigtriangledown}_{a}\rho_{m}+\frac{2}{3}\theta\tilde{\bigtriangledown}_{a}\theta-2\tilde{\bigtriangledown}_{a}\bigtriangledown^{2}\psi+2(\frac{1}{3}\theta^{2}-\rho_{m})\tilde{\bigtriangledown}_{a}\psi=-\frac{3Gf''}{2\theta}\tilde{\bigtriangledown}_{a}(\tilde{\bigtriangledown}_{a}\tilde{\bigtriangledown}_{b}G)\nonumber\\&&-\frac{1}{2}(Gf''-12\dot{G}f''')\tilde{\bigtriangledown}_{a}G+72f''\tilde{\bigtriangledown}_{a}\dot{G}+2(\frac{1}{2}(f'G-f)-24f''\dot{G})\tilde{\bigtriangledown}_{\psi},
 \label{eq81}
\end{eqnarray}
which is the second integrability condition in $f(G)$ gravity.
In the limits to the $\Lambda$CDM, it means $f(G)=G$, the right hand side of  Eq. \ref{eq83} vanishes and we obtain \cite{abebe2017integrability, sami2021covariant,maartens1998covariant}
\begin{equation}
 \tilde{\bigtriangledown}_{a}\rho_{m}+\frac{2}{3}\theta\tilde{\bigtriangledown}_{a}\theta-2\tilde{\bigtriangledown}_{a}\bigtriangledown^{2}\psi+2(\frac{1}{3}\theta^{2}-\rho_{m})\tilde{\bigtriangledown}_{a}\psi=0,
\end{equation} which coincides with $\Lambda$CDM.
\subsubsection{The peculiar velocity\\}
By taking the gradient of Eq. \ref{eq74} together with eq. \ref{eq44}, we can get the peculiar velocity and can be presented as follows
\begin{eqnarray}
 &&v_{a}=-\frac{1}{(\frac{1}{2}-\frac{3\dot{G}f''}{8\theta^{2}})\rho_{m}}\tilde{\bigtriangledown}_{a}\dot{\psi}+\frac{1}{(\frac{1}{2}-\frac{3\dot{G}f''}{8\theta^{2}})\rho_{m}}
 (\frac{Gf''}{2\theta}-\frac{G\dot{G}(f'')^{2}}{3})\tilde{\bigtriangledown}_{a}\dot{G}\nonumber\\&&+\frac{\dot{G}}{(\frac{1}{2}-\frac{3\dot{G}f''}{8\theta^{2}})\rho_{m}}\left(\frac{f''}{2\theta}+\frac{Gf''}{3}(f'''\dot{G}-\frac{\theta f''}{3})+\frac{Gf'''}{2\theta}\right)\tilde{\bigtriangledown}_{a}G.
 \label{eq83}
\end{eqnarray}
In the limit to $\Lambda$CDM, it means $f(G)=G$,  from Eq. \ref{eq83} we obtain \cite{abebe2017integrability, sami2021covariant,maartens1998covariant}
\begin{equation}
 v_{a}=-\frac{2}{\rho_{m}}\tilde{\bigtriangledown}_{a}\dot{\psi},
\end{equation} which coincides with the $\Lambda$CDM. This equation
  evolves as
\begin{equation}
 \dot{v}_{a}+\frac{1}{3}\theta v_{a}=-A_{a}
 \label{eq85}
\end{equation}
From equatios eq. \ref{eq79} and eq. \ref{eq85}, the peculiar velocity  and $4$-acceleration couple, and can be decoupled by taking the second derivative of eq. \ref{eq85} to make a second order differential equation in $v_{a}$ to produce
\begin{equation}
 \ddot{\mathcal{V}}_{a}+\theta \dot{\mathcal{V}}_{a}-\frac{1}{9}\mathcal{V}{a}=0.
 \label{eq86}
\end{equation}
From Eq. \ref{eq86}, the velocity perturbations is scale independent  since no spatial derivatives present. This equation is similar to the one obtained in \cite{maartens1998covariant} and its solutions were found and presented.
%\clearpage
\section{Linear perturbation equations in modified $f(G)$ gravity}\label{pert}
\subsection{First- and second- order evolution equations}
The vector covariant gradient variables can be defined as
\begin{eqnarray}
 D^{m}_{a}=a\frac{\tilde{\bigtriangledown}_{a}\rho_{m}}{\rho_{m}},
 Z_{a}=a\tilde{\bigtriangledown}_{a}\theta,
 \mathcal{G}_{a}=a\tilde{\bigtriangledown}_{a}G,
 \mathbf{G}_{a}=a\tilde{\bigtriangledown}_{a}\dot{G},
 \mathcal{A}_{a}=aA_{a},
 \mathcal{V}^{m}_{a}=av_{a},
 \label{eq88}
\end{eqnarray}
where $D^{m}_{a}$, $Z_{a}$, $\mathcal{G}_{a}$, $\mathbf{G}_{a}$, $\mathcal{A}_{a}$ and $\mathcal{V}^{m}_{a}$ represent the gradient variables for total matter fluid, the volume expansion, Gauss-Bonnet density fluid, Gauss-Bonnet momentum density, comoving acceleration and the velocity inhomogeneity of the matter respectively. In addition to the gradient variables defined in  our published paper \cite{munyeshyaka2022multifluid}, we included $\mathcal{A}_{a}$ and $
 \mathcal{V}^{m}_{a}$ to account for the  contribtion of velocity perturbations to the large scale structure formation.
We are interested in structure formation of the large scale structures. This  follows a spherical clustering mechanism for which the scalar parts  of the defined gradient variables play a key role. Therefore, we need to extract scalar parts of the perturbations by applying  a local decomposition scheme as
\begin{eqnarray}
 a \tilde{\bigtriangledown}_{a}X=\frac{1}{3}h_{ab}X+\varSigma^{x}_{ab}+X_{[ab]},
\end{eqnarray}
where $\varSigma^{x}_{ab}=X_{(ab)}-\frac{1}{3}h_{ab}X$ represents shear and $X_{[ab]}$ represents the vorticity which vanishes when extracting the scalar contribution.
By applying the comoving $a \tilde{\bigtriangledown}_{a}$ to the  vector gradient variables, Eq. \ref{eq88}, the scalar gradient variables can be represented as
\begin{eqnarray}
 &&\Delta_{m}=a\tilde{\bigtriangledown}_{a} D^{m}_{a}\;,
 Z=a\tilde{\bigtriangledown}_{a}Z_{a}\;,
 \mathcal{G}=a\tilde{\bigtriangledown}_{a}\mathcal{G}_{a}\;,
 \mathbf{G}=a\tilde{\bigtriangledown}_{a}\mathbf{G}_{a}\;,
 \mathcal{A}=a\tilde{\bigtriangledown}_{a}\mathcal{A}_{a}\;,\nonumber\\&&
\mathcal{V}^{m}=a\tilde{\bigtriangledown}_{a} \mathcal{V}^{m}_{a}\;.
\end{eqnarray}
 Starting with the time derivative of  the defined scalar parts of the gradient variables, we present the first- and the
second-order evolution equations responsible  for  the growth of matter and velocity perturbations. The system of equations governing the evolutions of these scalar fluctuations are given as follows
\begin{eqnarray}
 &&\dot{Z}=\left(-\frac{2}{3}\theta+\frac{2}{3}\theta^{2}\dot{G}f''\right)Z-\frac{\rho_{m}}{2}\Delta_{m}-\frac{1}{2}\left[(1-G)f'-f''+\frac{4}{9}\theta^{3}\dot{G}f'''\right]\mathcal{G}\nonumber\\
 && \quad\quad-\frac{2}{9}\theta^{3}f''\mathbf{G}+\tilde{\bigtriangledown}^{2} \mathcal{A}
 +\left[-\frac{1}{3}\theta^{2}-\frac{1}{2}\left(\rho_{m}+(f-Gf')+\frac{4}{9}\theta^{3}\dot{G}f''\right)\right]\mathcal{A}\;
 \label{eq91},\\
 && \dot{\Delta}_{m}+Z+\theta \mathcal{A}+\bigtriangledown^{2}V_{m}=0\;
 \label{eq92},\\
 && \dot{\mathcal{V}}_{m}+\mathcal{A}=0\;
 \label{eq93},\\
 && \dot{\mathcal{A}}+(\frac{1}{3}\theta-\frac{G\dot{G}}{6H}f'')\mathcal{A}+(\frac{1}{2}+\frac{3\dot{G}f''}{8\theta^{2}})\rho_{m}V_{m}
-\left(\frac{G\dot{G}(f'')^{2}}{3}+\frac{2}{9}\theta^{2}f''+\frac{Gf''}{2\theta}\right)\mathbf{G}\nonumber\\&&-\left[\left(f'''\dot{G}-\frac{\theta}{3}f''\right)\left(\frac{2}{9}\theta^{2}+\frac{G\dot{G}f''}{3} \right)+\frac{\dot{G}}{2\theta}\left(f''+Gf'''\right)\right]\mathcal{G}=0\;
\label{eq94},\\
&&\dot{\mathcal{G}}-\mathbf{G}-\dot{G}\mathcal{A}=0\;,
\label{eq95}\\
&&\dot{\mathbf{G}}-\frac{\dddot{G}}{\dot{G}}\mathcal{G}-\ddot{G}\mathcal{A}=0\;,
\label{eq96}
\end{eqnarray}
\begin{eqnarray}
&& \ddot{\Delta}_{m}-\left(-\frac{2}{3}\theta +\frac{2}{3}\theta^{2}\dot{G}f''\right)\dot{\Delta}_{m}-\frac{\rho_{m}}{2}\Delta_{m}-\Big[-\frac{5}{3}\theta^{2}-\left(\rho_{m}+(f-Gf')+\frac{4}{9}\theta^{3}\dot{G}f''\right)\nonumber\\&& -\frac{\theta \dot{G}^{2}(f'')^{2}}{3}+\frac{10}{9}\theta^{3}\dot{G}f''\Big]\dot{V}_{m}
  -\left(\frac{\theta}{2}+\frac{3G\dot{G}f''}{8\theta}\right)\rho_{m}V^{m}-\frac{2}{3}\theta^{2}\dot{G}f''\bigtriangledown^{2}V_{m}\nonumber \\
 && +\left[\frac{G\dot{G}\theta(f'')^{2}}{3}+\frac{Gf''}{2}-\frac{4}{9}\theta^{3}f''\right]\dot{\mathcal{G}}+\Big[\theta \left(\dot{G}f'''-\frac{\theta f''}{3}\right)\left(\frac{2}{9}\theta^{2}+\frac{G\dot{G}f''}{3}\right)\nonumber\\&&+\frac{\dot{G}}{2}(f''+Gf''')-\frac{1}{2}\left((1-G)f'-f''+\frac{4}{9}\theta^{3}\dot{G}f'''\right)\Big]\mathcal{G}=0\;,
 \label{eq97}\\
 && \ddot{V}^{m}+\left(\frac{1}{3}\theta+\frac{2}{9}\theta^{2}\dot{G}f''+\frac{G\dot{G}^{2}(f'')^{2}}{3}\right)\dot{V}_{m}-\left(\frac{1}{2}+\frac{3\dot{G}f''}{8\theta^{2}}\right)\rho_{m}V_{m}
+\Big[\frac{\dot{G}}{2\theta}\left(f''+Gf'''\right)\nonumber\\
 &&+\left(f'''\dot{G}-\frac{\theta}{3}f''\right)\left(\frac{2}{9}\theta^{2}+\frac{G\dot{G}f''}{3} \right)\Big]\mathcal{G} +\left(\frac{G\dot{G}(f'')^{2}}{3}+\frac{2}{9}\theta^{2}f''+\frac{Gf''}{2\theta}\right)\dot{\mathcal{G}}=0\;
\label{eq98}, \\
&& \ddot{\mathcal{G}}-\left(\frac{G\dot{G}^{2}(f'')^{2}}{3}+\frac{2}{9}\theta^{2}\dot{G}f''+\frac{G\dot{G}f''}{2\theta}\right)\dot{\mathcal{G}}-\Big[\frac{G\dot{G}^{3}(f'')^{2}}{3}+\frac{2}{9}\theta^{2}\dot{G}^{2}+\frac{G\dot{G}^{2}f''}{\theta}\nonumber\\&&-2\ddot{G}-\frac{1}{3}\theta \dot{G}\Big]\dot{V}_{m}+\left(\frac{1}{2}+\frac{3\dot{G}f''}{8\theta^{2}}\right)\dot{G}\rho_{m}V_{m}
-\Big[\dot{G}\left(f'''\dot{G}-\frac{\theta}{3}f''\right)\left(\frac{2}{9}\theta^{2}+\frac{G\dot{G}f''}{3} \right)\nonumber\\&&+\frac{\dot{G}^{2}}{2\theta}\left(f''+Gf'''\right)+\frac{\dddot{G}}{\dot{G}}\Big]\mathcal{G}
=0\; .
\label{eq99}
\end{eqnarray}
\subsection{Harmonic decomposition}
The above  evolution equations form a coupled system of harmonic oscillator differential equations of the form \cite{abebe2012covariant,ntahompagaze2018study,abebe2015breaking,carloni2006gauge,murorunkwere20211+,munyeshyaka2021cosmological}:
\begin{equation}
 \ddot{X}+A\dot{X}+BX=C(Y,\dot{Y}),
 \label{eq100}
\end{equation}
where $A$, $B$ and $C$ are independent of $X$ and they represent friction (damping), restoring and
source forcing terms respectively. To solve Eq. \ref{eq100}, a separation of variables technique is applied such that
\begin{eqnarray}
&&X(x,t)=X(\vec{x}).X(t)\nonumber\\
&& Y(x,t)=Y(\vec{x}).Y(t). \nonumber
\end{eqnarray}
 The  evolution equations Eq. \ref{eq91} through to
 Eq. \ref{eq99} complicate to be solved with this technique, so we apply a harmonic decomposition approach to these equations by using the eigenfunctions and their corresponding wave-number. Starting with
 \begin{eqnarray}
  && X=\sum_{k}X^{k}(t).Q_{k}(\vec{x}),\nonumber \\
  && Y=\sum_{k}Y^{k}(t).Q_{k}(\vec{x}), \nonumber
 \end{eqnarray}
 and with little algebra,  we present the evolution equations as
\begin{eqnarray}
 && \ddot{\Delta}^{k}_{m}-\left(-\frac{2}{3}\theta +\frac{2}{3}\theta^{2}\dot{G}f''\right)\dot{\Delta}^{k}_{m}-\frac{\rho_{m}}{2}\Delta^{k}_{m}-\Big[-\frac{5}{3}\theta^{2}-\left(\rho_{m}+(f-Gf')+\frac{4}{9}\theta^{3}\dot{G}f''\right)\nonumber \\
 &&-\frac{\theta \dot{G}^{2}(f'')^{2}}{3}+\frac{10}{9}\theta^{3}\dot{G}f''\Big]\dot{V}^{k}_{m}-\left(\frac{\theta}{2}\rho_{m}+\frac{3G\dot{G}f''\rho_{m}}{8\theta}-\frac{2}{3}\theta^{2}\dot{G}f''\frac{k^{2}}{a^{2}}\right)V_{m}^{k}\nonumber\\&&+\left[\frac{G\dot{G}\theta(f'')^{2}}{3}+\frac{Gf''}{2}-\frac{4}{9}\theta^{3}f''\right]\dot{\mathcal{G}}^{k}+\Big[\frac{\dot{G}}{2}(f''+Gf''')+\theta \left(\dot{G}f'''-\frac{\theta f''}{3}\right)\left(\frac{2}{9}\theta^{2}+\frac{G\dot{G}f''}{3}\right)\nonumber\\&&-\frac{1}{2}\Big((1-G)f'-f''+\frac{4}{9}\theta^{3}\dot{G}f'''\Big)\Big]\mathcal{G}^{k}=0\;\label{eq101}, \\
 && \ddot{V}_{m}^{k}+\left(\frac{1}{3}\theta+\frac{2}{9}\theta^{2}\dot{G}f''+\frac{G\dot{G}^{2}(f'')^{2}}{3}\right)\dot{V}^{k}_{m}-\left(\frac{1}{2}+\frac{3\dot{G}f''}{8\theta^{2}}\right)\rho_{m}V^{k}_{m}
+\Big[\frac{\dot{G}}{2\theta}\left(f''+Gf'''\right)\nonumber\\&&+\left(f'''\dot{G}-\frac{\theta}{3}f''\right)\left(\frac{2}{9}\theta^{2}+\frac{G\dot{G}f''}{3} \right)\Big]\mathcal{G}^{k}+\left(\frac{G\dot{G}(f'')^{2}}{3}+\frac{2}{9}\theta^{2}f''+\frac{Gf''}{2\theta}\right)\dot{\mathcal{G}}^{k}=0\;
\label{eq102}, \\
&& \ddot{\mathcal{G}}^{k}-\left(\frac{G\dot{G}^{2}(f'')^{2}}{3}+\frac{2}{9}\theta^{2}\dot{G}f''+\frac{G\dot{G}f''}{2\theta}\right)\dot{\mathcal{G}}^{k}-\left[\frac{G\dot{G}^{3}(f'')^{2}}{3}+\frac{2}{9}\theta^{2}\dot{G}^{2}+\frac{G\dot{G}^{2}f''}{\theta}-2\ddot{G}-\frac{1}{3}\theta \dot{G}\right]\dot{V}^{k}_{m}\nonumber \\
&& +\left(\frac{1}{2}+\frac{3\dot{G}f''}{8\theta^{2}}\right)\dot{G}\rho_{m}V^{k}_{m}
-\Big[\dot{G}\left(f'''\dot{G}-\frac{\theta}{3}f''\right)\left(\frac{2}{9}\theta^{2}+\frac{G\dot{G}f''}{3} \right)+\frac{\dot{G}^{2}}{2\theta}\left(f''+Gf'''\right)\nonumber\\&&+\frac{\dddot{G}}{\dot{G}}\Big]\mathcal{G}^{k}
=0\;.
\label{eq103}
\end{eqnarray}
where
\begin{eqnarray}
 &&\tilde{\bigtriangledown}^{2}=-\frac{k^{2}}{a^{2}}Q^{k},~ \dot{Q}_{k}(\vec{x})=0,~ k=\frac{2 \pi a}{\lambda},
\end{eqnarray}
with $Q^{k}$, $k$ and $\lambda$ are the eigenfunction of the comoving spatial Laplacian, the order of harmonic(wave-number) and the physical wavelength of the mode respectively.
We analyse the linear evolution equations of the matter energy-density and velocity perturbations with cosmological redshift and find both analytical and numerical solutions.
\subsection{Redshift transformation}
We  transform  the linear perturbation equations from time derivative  into redshift derivative \cite{sahlu2020scalar,sami2021covariant,ntahompagaze2018study,murorunkwere20211+,munyeshyaka2021cosmological,munyeshyaka2022multifluid,venikoudis2022late}. Starting with
\begin{eqnarray}
        && a=\frac{a_{0}}{1+z},\nonumber\\
        &&\dot{f}=-(1+z)Hf',\nonumber\\
        &&\ddot{f}=(1+z)^{2}H\Big[H'f'+Hf''\Big]+(1+z)H^{2}f',
       \end{eqnarray}
        where $'$ means differentiation with respect to redshift. Throughout this manuscript, $a_{0}$ is set to $1$.
  Eq. \ref{eq101}- \ref{eq103} can be written into redshift space as follow:
\begin{eqnarray}
 && \Delta''_{m}-\frac{1}{(1+z)}\left(\frac{1}{2}+\mathcal{Y}\right)\Delta'_{m}-\frac{3\Omega_{m}}{2(1+z)^{2}}\Delta_{m}+\frac{1}{1+z}\left[3H\left(-1-\Omega_{m}+2\mathcal{X}\right)-3\mu \right]V'_{m}
 \nonumber \\
 && \quad \quad -\frac{1}{(1+z)^{2}}\left[3\Omega_{m}\left(\frac{3}{2}H+\frac{G\mathcal{Y}}{48H^{2}}\right)-\frac{\mathcal{Y}}{H} \frac{k^{2}}{a^{2}}\right]V_{m}-\frac{3}{(1+z)}\left(\varepsilon-9H^{2}f''\right)\mathcal{G}'\nonumber \\
 && \quad \quad+\frac{\beta}{(1+z)^{2}H^{2}}\mathcal{G}=0\;\label{eq105}, \\
 && V''_{m}+\frac{1}{(1+z)}\left(\frac{1}{2}-\nu \right)V'_{m}-\frac{3\Omega_{m}}{2(1+z)^{2}}\left(1+\frac{\mathcal{Y}}{8H\theta^{2}}\right)V_{m}
+\frac{\eta}{(1+z)^{2}H^{2}}\mathcal{G}\nonumber\\
&& \quad \quad-\frac{\varepsilon}{(1+z)H}\mathcal{G}'=0\;
\label{eq106},
\end{eqnarray}
\begin{eqnarray}
&& \mathcal{G}''-\frac{1}{(1+z)H}\left(\zeta -\frac{5\theta}{6}+\frac{G\mathcal{Y}}{36H^{2}}\right)\mathcal{G}' +\frac{1}{(1+z)H}\left[\dot{G}\left(\dot{G}\varepsilon+\frac{G\mathcal{Y}}{36H^{2}}-\frac{1}{3}\right)-2\ddot{G}\right]V'_{m}\nonumber \\
&& \quad \quad+\frac{3\Omega_{m}}{2(1+z)^{2}}\left(1+\frac{\mathcal{Y}}{8H\theta^{2}}\right)\dot{G}V_{m}
-\frac{\gamma}{(1+z)^{2}H^{2}}\mathcal{G}
=0\; .
\label{eq107}
\end{eqnarray}
 These equations (Eq. \ref{eq105}- \ref{eq107}) differ from the ones obtained in the work done in \cite{munyeshyaka2022multifluid}   because  both energy density perturbations and perturbations due to Gauss-Bonnet fluid couple with the velocity perturbations. In the limit to $\Lambda$CDM, it means $f(G)=G$, from Eq. \ref{eq105}- \ref{eq107} we obtain
\begin{eqnarray}
&& \Delta''_{m}-\frac{1}{2(1+z)}\Delta'_{m}-\frac{3\Omega_{m}}{2(1+z)^{2}}\Delta_{m}-\frac{3H}{1+z}\left(1+\Omega_{m}\right)V'_{m}
  -\frac{9H\Omega_{m}}{2(1+z)^{2}}V_{m}=0\;,
 \label{eq108}\\
 && V''_{m}+\frac{1}{2(1+z)}V'_{m}-\frac{3\Omega_{m}}{2(1+z)^{2}}V_{m} =0\;
 \label{eq109}\\
 && \mathcal{G}''=0,
 \end{eqnarray} which coincides with $\Lambda$ CDM.
These results are related to the ones obtained in the works presented in \cite{sami2021covariant,abebe2017integrability}.
When it comes to Eq. \ref{eq108}, matter energy density couple with velocity perturbations, and will be decoupled in finding analytical solutions using solutions of Eq. \ref{eq109}.
\subsection{ Analytical and numerical solutions\\}
\begin{figure}[h!]
\begin{minipage}{0.45 \textwidth}
   \includegraphics[width=55mm, height=55mm]{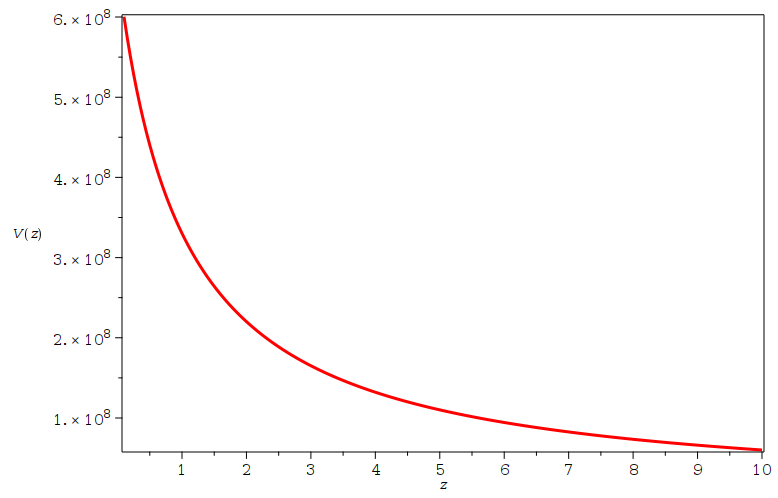}
  \caption{Plot of velocity perturbations versus redshift for $\Lambda$CDM limits, Eq. \ref{eq109}. Normalization was made by multiplying $10^{2}$ to the $V(z)$.}
  \label{Fig1}
  \end{minipage}
  \begin{minipage}{0.5\textwidth}
     \includegraphics[width=55mm, height=55mm]{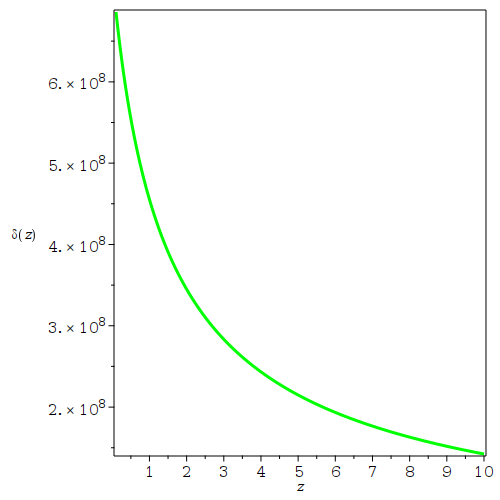}
  \caption{Plot of energy density perturbations versus redshift for $\Lambda$CDM limits, Eq. \ref{eq108}. Normalization was made by multiplying $10^{2}$ to the $\delta(z)$.}
  \label{Fig2}
  \end{minipage}
  \end{figure}

\begin{figure}
     \begin{minipage}{0.45\textwidth}
  \includegraphics[width=55mm]{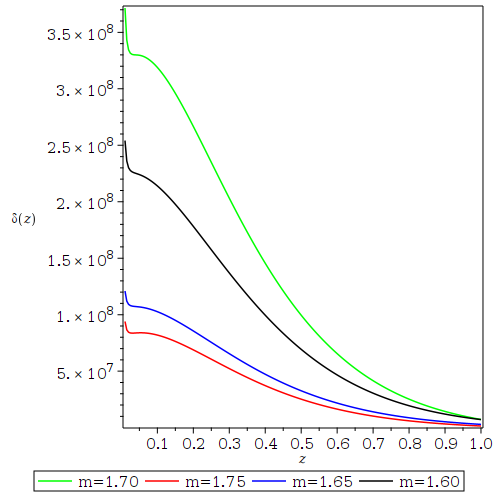}
  \caption{Plot of energy density perturbations versus redshift of Eq. \ref{eq105}-\ref{eq107} for different values of $m$ with $k\ll 1$. To find numerical results, we considered the relation $\frac{k^{2}}{a^{2}H^{2}}=\frac{16\pi^{2}}{\lambda^{2}(1+z)^{2}}$ throughout all plots.}
  \label{Fig3}
  \end{minipage}
  \begin{minipage}{0.5\textwidth}
  \includegraphics[width=55mm]{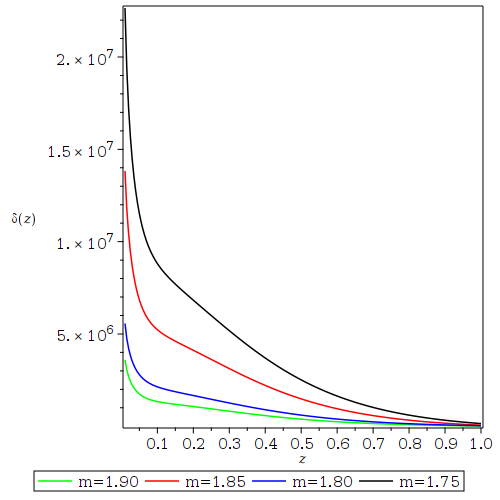}
  \caption{Plot of energy density perturbations versus redshift of Eq. \ref{eq105}-\ref{eq107} for different values of $m$ with $k\gg 1$.}
  \label{Fig4}
  \end{minipage}
 \begin{minipage}{0.45\textwidth}
  \includegraphics[width=55mm]{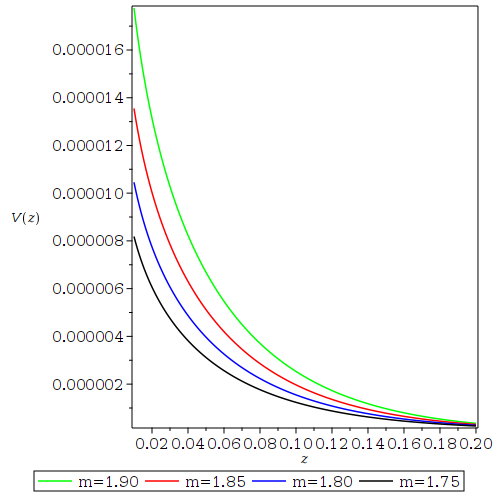}
  \caption{Plot of velocity perturbations versus redshift of Eq. \ref{eq105}-\ref{eq107} for different values of $m$ with $k\ll 1$. }
  \label{Fig5}
 \end{minipage}
 \begin{minipage}{0.5\textwidth}
  \includegraphics[width=55mm]{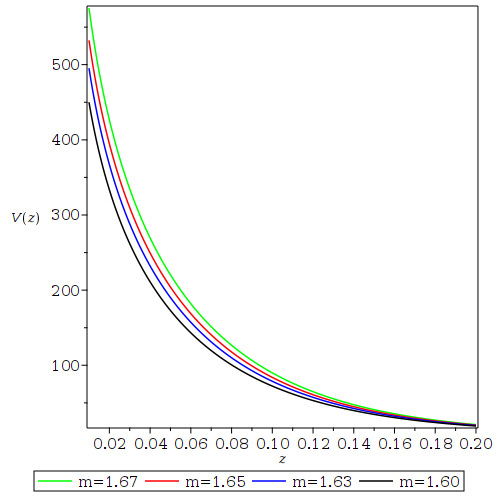}
  \caption{Plot of velocity perturbations versus redshift of Eq. \ref{eq105}-\ref{eq107} for different values of $m$ with $k\gg 1$.}
  \label{Fig6}
  \end{minipage}
 \end{figure}

We define the normalised energy density contrast for matter fluid as
\begin{equation*}
 \delta(z)=\frac{\Delta^{k}_{m}(z)}{\Delta(z_{0})},
\end{equation*}
where $\Delta(z_{0})$ is the matter energy density at the initial redshift.
The normalized velocity can be defined as
\begin{equation*}
 \mathcal{V}(z)=\frac{V(z)}{V(z_{0})}
\end{equation*}
We need to find analytical solutions of Eq. \ref{eq108} and Eq. \ref{eq109}.
The second order perturbation equation Eq. \ref{eq108} forms an non-closed system while Eq. \ref{eq109} forms a closed system therefore gives an easy task in finding analytical solutions. The analytical solution of Eq. \ref{eq109} is presented as
\begin{eqnarray}
 && V(z)=C_{1}(1+z)^{\frac{1}{4}+\frac{\sqrt{205}}{20}}+C_{2}(1+z)^{\frac{1}{4}-\frac{\sqrt{205}}{20}}\;.
 \end{eqnarray}
 Since  we get the analytical solutions of the velocity perturbations, we can use them to find the solutions for energy density perturbations (Eq. \ref{eq108}) and can be written as the following
 \begin{eqnarray}
 && \Delta(z)=C_{3}(1+z)^{(\frac{3}{4}-\frac{9\sqrt{5}}{20})}+C_{4}(1+z)^{(\frac{3}{4}+\frac{9\sqrt{5}}{20})}+C_{5}(1+z)^{(\frac{7}{4}-\frac{\sqrt{205}}{20})}\nonumber\\&&+C_{6}(1+z)^{(\frac{7}{4}+\frac{\sqrt{205}}{20})}\;,
\end{eqnarray}
 with $C_{5}=\frac{78-5\sqrt{205}}{100}C_{2}$ and $C_{6}=\frac{78+5\sqrt{205}}{100}C_{1}$, where $C_{1}$ through to $C_{6}$ are the constants which can be found by applying the initial conditions. We set the current value of the matter density parameter $\Omega_{m}\approx0.3$ based on the latest Planck data \cite{aghanim2020planck}, $V_{in}=10^{-5}$, $\dot{V}_{in}=0$, $\Delta_{in}=10^{-5}$ and $\dot{\Delta}_{in}=0$.
The numerical solutions of energy density and velocity perturbations for  GR limits (Eq. \ref{eq108} and Eq. \ref{eq109}) are presented in Fig. \ref{Fig1} and Fig. \ref{Fig2} respectively. We set the initial condition $z_{0}=1100$ during the  matter-radiation decoupling era and we explore the features of fractional energy density and velocity perturbations $\delta(z)$ and $\mathcal{V}(z)$ with redshift range $0\le z\le 1100$. For pedagogical purpose, we considered a polynomial $f(G)$ model represented as \cite{goheer2009coexistence,uddin2009cosmological,rastkar2012phantom,munyeshyaka2022multifluid}
\begin{equation}
 f(G)=G-\frac{1}{2}\left(\sqrt{\frac{6m(m-1)G}{(m+1)^{2}}}+AG^{\frac{3}{4}m(1+w)}\right),
 \label{eq37}
\end{equation} where
\begin{equation*}
 A=\frac{8\rho_{0}(m-1)\left[13824m^{9}(m-1)^{3}\right]^{-\frac{1}{4}m(1+w)}}{4+m\left[3m(1+w)(w+\frac{4}{3})-18w-19\right]},
 \label{eq27}
\end{equation*}
where $m$ is a positive constant, to  find numerical solutions of Eq. \ref{eq105}- \ref{eq107}. For $m=1$, we get $f(G)=G$ and we  recover GR limits.
  For  $m> 1$ and   $4+m\left[3m(1+w)(w+\frac{4}{3})-18w-19\right]\neq 0$  this $f(G)$ model produces an accelerated phase of the Universe.
 \cite{cognola2006dark} discussed the cases where  Einstein or Gauss-Bonnet term dominates one another for this $f(G)$ model. We chose this $f(G)$ model for a quantitative analysis of the derived perturbation equations Eq. \ref{eq105} - \ref{eq107}, it is a  viable model which is  compatible with cosmological observations and  can be treated among the representative examples of models that could account for the late-time acceleration of the universe without the need for dark energy \cite{goheer2009coexistence, de2009construction}. The Hubble parameter in redshift space  for a dust dominated universe ($w=0$) is presented as $H(z)=\frac{2m}{3}(1+z)^{\frac{3}{2m}}$, for GR limits, we set $m=1$. The Gauss-Bonnet term is given by $ G=24H^{2}\left(H^{2}+\dot{H}\right)$ which in redshift space is presented as $G=\frac{64}{9}m^{3}\Big(\frac{2m}{3}-1\Big)(1+z)^{\frac{6}{m}}$ and the energy density parameter $\rho_{m}=3H^{2}-\frac{1}{2}\Big(Gf'-f-24\dot{G}H^{3f''}\Big)$. To get $\rho_{m}$ in redshift space, we replace $H$, $G$ and $f$ by their expressions. We assume that the dynamics of the universe is driven by  the power-law scale factor of the form $a(t)=t^{\frac{2m}{3(1+w)}}$ and we consider $H=\frac{\dot{a}}{a}$. For simplicity we set $f(G)$ to $f$  and considered the initial conditions $V_{in}=10^{-5}$, $\dot{V}_{in}=0$, $\Delta_{in}=10^{-5}$ and $\dot{\Delta}_{in}=0$, $\mathcal{G}_{in}=10^{-5}$ and $\dot{\mathcal{G}}_{in}=0$ to find numerical solutions of Eq. \ref{eq105}- \ref{eq107} which are  presented  in Fig. \ref{Fig3}-\ref{Fig6} for both short( for example $\lambda=0.1Mpc, 0.001Mpc, 0.001Mpc$)- and long($\lambda=10Mpc, 100Mpc$)- wavelength modes.
 We analysed the behavior of velocity and the energy density contrast for different values of $m$. For Velocity perturbations, $V(z)$, the range of the parameter $m$ which presented good results is for $1.75\leq m\leq 1.90$ for long wavelength modes and $1.60\leq m\leq 1.70$ for short wavelength modes.\\
 For the energy density contrast $\delta(z)$, we considered $1.60 \leq m \leq 1.70$ for long wavelength modes and $1.75\leq m\leq 1.90$ for short wavelength modes. From all considered ranges of $m$, both energy density contrast and velocity perturbations decay with increase in redshift. The decay of energy density contrast seems to be non linear compared with the $\Lambda$CDM limits. \\For the velocity perturbations, the decay mimics  the $\Lambda$CDM limits, only differs for higher amplitudes. For both the energy density contrast and velocity perturbations, the numerical results present higher amplitudes for short wavelength than for long wavelength modes.
 \clearpage
 \subsection{Quasi-static aproximation}
\begin{figure}
\begin{minipage}{0.5\textwidth}
  \includegraphics[width=55mm,height=55mm]{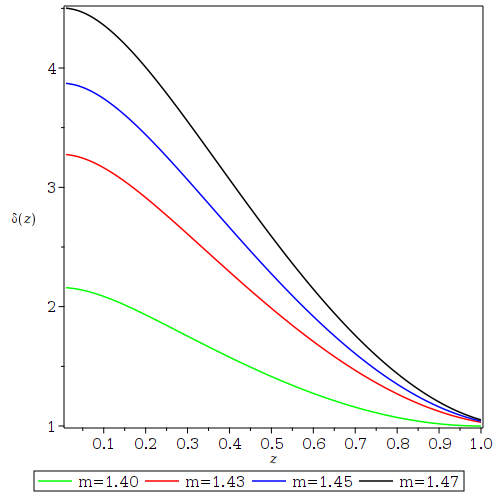}
  \caption{Plot of energy density perturbations versus redshift of Eq. \ref{eq112}-\ref{eq113} for different values of $m$ with $k\ll 1$.}
  \label{Fig7}
 \end{minipage}
 \begin{minipage}{0.45\textwidth}
   \includegraphics[width=55mm,height=55mm]{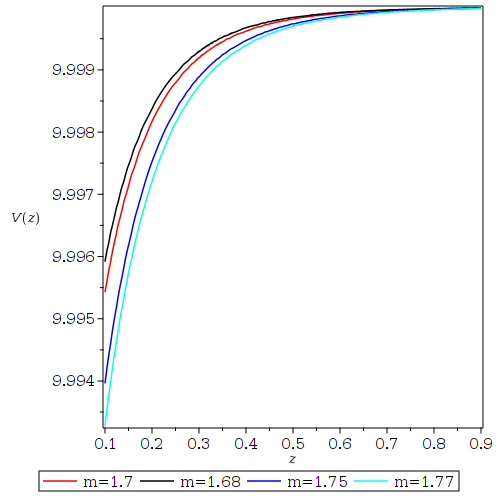}
  \caption{Plot of velocity perturbations versus redshift of Eq. \ref{eq112}-\ref{eq113} for different values of $m$ with $k\ll 1$.}
  \label{Fig8}
 \end{minipage}
   \begin{minipage}{0.5\textwidth}
  \includegraphics[width=80mm,height=50mm]{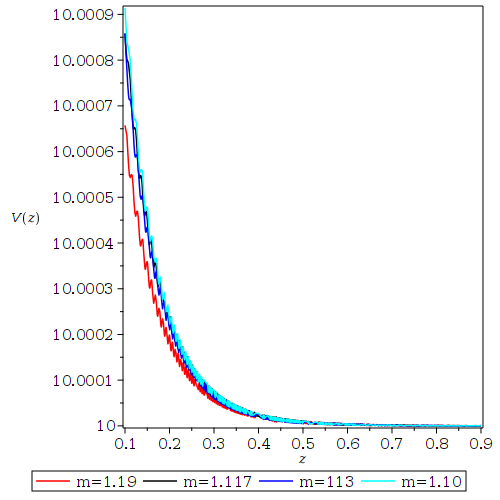}
  \caption{Plot of velocity perturbations versus redshift of Eq. \ref{eq112}-\ref{eq113} for different values of $m$ with $k\ll 1$.}
  \label{Fig9}
  \end{minipage}
 \end{figure}

  For small scale perturbations, perturbations due to Gauss-Bonnet contributions will slowly evolve compared to the matter energy density perturbations\cite{abebe2012covariant,abebe2017integrability,sahlu2020scalar,sami2021covariant} therefore, we use a Quasi-static approximation, where the temporal derivatives  in the Gauss-Bonnet term $G$ are neglected it means $\ddot{\mathcal{G}}\approx 0$ and $\dot{\mathcal{G}}\approx 0$. Therefore the overall dynamics of the density and velocity perturbations of the system of equations Eq. (\ref{eq101})-(\ref{eq103})  after redshift transformation can be presented as
\begin{eqnarray}
&& \Delta''_{m}-\frac{1}{(1+z)}\left(\frac{1}{2}+\mathcal{Y}\right)\Delta'_{m}-\frac{3\Omega_{m}}{2(1+z)^{2}}\Delta_{m}\nonumber \\
&&\quad\quad+\frac{1}{1+z}\left[\frac{\beta}{H\gamma}\left[\dot{G}\left(\dot{G}\varepsilon+\frac{G\mathcal{Y}}{36H^{2}}-\frac{1}{3}\right)-2\ddot{G}\right]+3H\left(-1-\Omega_{m}+2\mathcal{X}\right)-3\mu \right]V'_{m}
 \nonumber \\
 && \quad \quad -\frac{1}{(1+z)^{2}}\left[3\Omega_{m}\left(-\frac{\beta}{2\gamma}\left(1+\frac{\mathcal{Y}}{8H\theta^{2}}\right)\dot{G}+\frac{3}{2}H+\frac{G\mathcal{Y}}{48H^{2}}\right)-\frac{\mathcal{Y}}{H} \frac{k^{2}}{a^{2}}\right]V_{m} =0\;\label{eq112}, \\
 && V''_{m}+\frac{1}{(1+z)}\left(\frac{1}{2}-\nu-\frac{\eta}{H\gamma}\dot{G}\left(\dot{G}\varepsilon+\frac{G\mathcal{Y}}{36H^{2}}-\frac{1}{3}\right)-2\ddot{G}\right)V'_{m}\nonumber\\
 &&\quad\quad-\frac{3\Omega_{m}}{2(1+z)^{2}}\left[\left(1+\frac{\mathcal{Y}}{8H\theta^{2}}\right)\left(1-\frac{\eta}{\gamma}\right)\right]V_{m}
=0\;.\label{eq113}
\end{eqnarray}
From Eq. \ref{eq112}, energy density couple with velocity perturbations while from Eq. \ref{eq113} energy density perturbations decouple from velocity perturbations. Numerical solutions of Eq. \ref{eq112} and Eq. \ref{eq113} are presented in Fig. \ref{Fig7}-\ref{Fig9}.
 In order to get the numerical results of Eq. \ref{eq112} and Eq. \ref{eq113} simultaneously, we set different values of the parameter $m$, the consideration of the initial conditions and the normalization. The numerical solutions for both energy density contrast and velocity perturbations were only evaluated for long wavelength modes. For $1.40 \leq m\leq1.47$, energy density contrast decays with increasing redshift as presented  in fig. \ref{Fig7}. For $1.10\leq m \leq1.19$, the velocity perturbations decay with increase in redshift as presented in fig. \ref{Fig8}, which is not the case for the consideration of $1.68\leq m\leq 1.77$ as presented in fig. \ref{Fig9}.  We depict unrealistic behavior of the velocity perturbations. For both the energy density contrast and the velocity perturbations, the $\Lambda$CDM limit is recovered for the case $f(G)=G$ or $m=1$.
 \clearpage
\section{Discussion and Conclusion}\label
{concsec}
 In this paper, we have considered a specific $f(G)$ gravity model which provides the expansion history of the Universe for a power-law cosmological scale factor which can mimic the $\Lambda CDM$ model. The considered non-linear $f (G)$ model has been studied to be able to drive the inflationary era in the early epoch and can describe the late time cosmic acceleration. For example, in Capozziello et al \cite{capozziello2014noether}, the considered $f(R,G)$ model with both $R$ and $G$ treated to be non-linear. This paper presents a detailed analysis of the $1+3$ covariant velocity and energy density  perturbations of the quasi-newtonian cosmology in the context of $f(G)$ gravity. From propagation and constraints equations, we have derived first and second integrability conditions for $f(G)$ gravity, where we were able to obtain covariant velocity, acceleration and the modified Poisson equations for both $f(G)$ gravity and $\Lambda$CDM model. The obtained velocity and acceleration equations were coupled then decoupled to  a second order differential equations. Since our interest extends to the  large-scale structure formation, we have derived scalar velocity and energy density perturbation equations. In doing so, we have defined gauge invariant scalar gradient variables  and derived their corresponding scalar linear evolution equations. We then  apply harmonic decomposition method together with the redshift transformation technique to make equations manageable for analysis. From that, we solve the whole system of equations without considering the quasi-static approximation. We obtain both of the analytical  and numerical solutions for both velocity and matter energy density for $\Lambda$CDM limits while numerical results were obtained for the considered $f(G)$ model. We computed the growth of fractional energy density $\delta(z)$ and velocity $\mathcal{V}(z)$ for the considered $f(G)$ model  and the power-law cosmological scale factor. From the plots, We depict that both $\delta(z)$ and $V(z)$ decay with redshift for long and short wavelength modes. For the case $m=1$, this decay of $\delta(z)$ and $V(z)$ with redshift  presents a feature mimicking the $\Lambda$CDM limits. \\ \\
On the other hand, we have considered the quasi-static approximation for analysing small scale perturbations where temporal derivatives of the Gauss-Bonnet fluid energy density and momentum were discarded. We show that $\delta(z)$ and $\mathcal{V}(z)$ decay with redshift for the considered range of parameter $m$.\\ \\ Some of the specific highlights of the present paper include:
\begin{itemize}
\item We have presented the integrability conditions, covariant modified Poisson equation in $f(G)$ gravity, evolution of velocity and  $4$-acceleration in modified $f(G)$ gravity in  Eq. \ref{eq73}, Eq. \ref{eq75},Eq. \ref{eq78}, Eq. \ref{eq81},Eq. \ref{eq83},  and energy density and velocity perturbations Eq. \ref{eq105} through to Eq. \ref{eq107}in quasi-newtonian space-time in $f(G)$ gravity which can be reduced to GR limits for the case of linear $f(G)$ gravity.
 \item During the analysis stage, we have considered short and long wavelength modes for the perturbation equations in a dust-Gauss-Bonnet fluids and considered a combination of non-linear $f(G)$ models with a linear $f(R)$ model such that, for the case $f(G)$  is choosen to be linear ($f(G)=G$), we recover GR limits.
 \item The numerical results of the velocity and energy density perturbations are  presented in figures Fig. \ref{Fig1}- \ref{Fig9} for both short and long wavelength modes. From the plots, for a linear $f(G)$ case, we depict the contribution of dust fluid in quasi-newtonian spacetime  universe for matter energy density and velocity perturbations which decay with increasing redshift.
 \item We presented the range of parameter $m$ for which both the energy density and velocity perturabtions decay with increase in redshift \cite{ ananda2009structure, abebe2013large,munyeshyaka2022multifluid}.
 \item The current results show that for both energy density and velocity perturbations, the quasi-newtonian spacetime in modified $f(G)$ gravity offers an alternative for large scale structure formation since the energy density and velocity perturbations decay with redshift  hence can provide a room  in the understanding of cosmic acceleration scenario.
\end{itemize}
In Quasi-Newtonian space-time without considering the quasi-static approximation, both energy density and velocity perturbations couple with the perturbations due to the Gauss-Bonnet contributions whereas by considering the quasi-static approximations, Both energy density and velocity perturbations decouple from the perturbations due to the Gauss-Bonnet contribution. By refering to the numerical results, the quasi-static approximation is not applicable for short-wavelength modes and for long-wavelength modes as $m$ get large  since the velocity perturbations do not decay with redshift for the considered range of parameter $m$.
In conclusion, we deduce that the energy density contrast and velocity perturbations decay with redshift for the considered $f(G)$ model  without quasi-static approximation for all range of parameter $m$ considered for both long and short wavelength modes. By considering quasi-static approximations, the energy density contrast decay with increase in redshift in long wavelength modes and the velocity perturbations decay with increasing redshift for $m$ getting closer to $1$, but show unrealistic features for large values of parameter $m$ in the long wavelength modes. Both energy density contrast and velocity perturbations results coincide with the $\Lambda$CDM limits for the case $f(G)=G$.  The  $\delta(z)$ and $\mathcal{V}(z)$ for the considered $f(G)$ model are consistent with the $\Lambda$CDM predictions for the considered range of parameter $m$, therefore the large-scale structure formation is enhanced.  The future work should consider different $f(G)$ models to check for the consistency with different observational predictions.
\section*{Acknowledgements}
We thank the anonymous reviewer(s) for the constructive comments towards the
signiﬁcant improvement of this manuscript. AM acknowledges that this work is supported  by the Swedish International Development Agency (SIDA)  to the International Science Program (ISP) through East African Astrophysics Research Network (EAARN) (grant number  AFRO:$05$).  AM also acknowledges the hospitality of the Department of Physics of the University of Rwanda, where this work was conceptualized and completed. AM  acknowledges  useful help from Both Dr. Heba Sami and Prof. Amare Abebe during the derivation of different equations. JN and MM acknowledge the ﬁnancial support provided by (SIDA) through to (ISP) to the University of
Rwanda via Rwanda Astrophysics, Space and Climate Science Research Group
(RASCSRG) grant number:RWA$01$.
\appendix
\section{Useful Linearised Differential Identities}
For all scalars $f$, vectors $V_a$ and tensors that vanish in the background,
$S_{ab}=S_{\la ab\ra}$, the following linearised identities hold:
\begin{eqnarray}
\left(\D_{\la a}\D_{b\ra}f\right)^{.}&=&\D_{\la a}\D_{b\ra}\dot{f}-\sfrac{2}{3}\Theta\D_{\la a}\D_{b\ra}f+\dot{f}\D_{\la a}A_{b\ra}\label{a0}\;,\\
\ep^{abc}\D_b \D_cf &=& 0 \label{a1}\;, \\
\ep_{cda}\D^{c}\D_{\la b}\D^{d\ra}f&=&\ep_{cda}\D^{c}\D_{( b}\D^{d)}f=\ep_{cda}\D^{c}\D_{ b}\D^{d}f=0\label{a2}\;,\\
\D^2\left(\D_af\right) &=&\D_a\left(\D^2f\right)
+\sfrac{1}{3}\tl{R}\D_a f \label{a4}\;,\\
\left(\D_af\right)^{\rd} &=& \D_a\dot{f}-\sfrac{1}{3}\Theta\D_af+\dot{f}A_a
\label{a14}\;,\\
\left(\D_aS_{b\cdots}\right)^{\rd} &=& \D_a\dot{S}_{b\cdots}
-\sfrac{1}{3}\Theta\D_aS_{b\cdots}
\label{a15}\;,\\
\left(\D^2 f\right)^{\rd} &=& \D^2\dot{f}-\sfrac{2}{3}\Theta\D^2 f
+\dot{f}\D^a A_a \label{a21}\;,\\
\D_{[a}\D_{b]}V_c &=&
-\sfrac{1}{6}\tl{R}V_{[a}h_{b]c} \label{a16}\;,\\
\D_{[a}\D_{b]}S^{cd} &=& -\sfrac{1}{3}\tl{R}S_{[a}{}^{(c}h_{b]}{}^{d)} \label{a17}\;,\\
\D^a\left(\ep_{abc}\D^bV^c\right) &=& 0 \label{a20}\;,\\
\label{divcurl}\D_b\left(\ep^{cd\la a}\D_c S^{b\ra}_d\right) &=& {\ts{1\over2}}\ep^{abc}\D_b \left(\D_d S^d_c\right)\;,\\
\text{curlcurl} V_{a}&=&\D_{a}\left(\D^{b}V_{b}\right)-\D^{2}V_{a}+\sfrac{1}{3}\tl{R}V_{a}\label{curlcurla}\;,
\label{a21}
\end{eqnarray}
\clearpage
\section{Used equations}
For more simplicity, we introduce here some quantities such as:
\begin{eqnarray}
 &&\quad \quad \alpha=(1-G)f'-f''+\frac{4}{9}\theta^{3}\dot{G}f'''\;, \nonumber \\
 && \quad \quad \beta=\frac{\dot{G}}{2}(f''+Gf''')+\theta \left(\dot{G}f'''-\frac{\theta f''}{3}\right)\left(\frac{2}{9}\theta^{2}+\frac{G\dot{G}f''}{3}\right)\nonumber \\
 && \quad\quad-\frac{1}{2}\left((1-G)f'-f''+\frac{4}{9}\theta^{3}\dot{G}f'''\right)\;, \nonumber\\
 && \quad\quad \gamma=\dot{G}\left(f'''\dot{G}-\frac{\theta}{3}f''\right)\left(\frac{2}{9}\theta^{2}+\frac{G\dot{G}f''}{3} \right)+\frac{\dot{G}^{2}}{2\theta}\left(f''+Gf'''\right)+\frac{\dddot{G}}{\dot{G}}\;, \nonumber \\
 && \quad \quad \eta=\left(f'''\dot{G}-\frac{\theta}{3}f''\right)\left(\frac{2}{9}\theta^{2}+\frac{G\dot{G}f''}{3} \right)+\frac{\dot{G}}{2\theta}\left(f''+Gf'''\right)\;,  \nonumber \\
 && \quad \quad \zeta =\frac{1}{3}\theta+\frac{G\dot{G}^{2}(f'')^{2}}{3}+\frac{2}{9}\dot{G}\theta^{2}f''\;, \nonumber \\
 && \quad\quad \varepsilon=\frac{G\dot{G}(f'')^{2}}{3}+\frac{2}{9}\theta^{2}f''+\frac{Gf''}{2\theta}\;, \nonumber \\
 && \quad\quad
 \mu =\frac{4}{3}\theta+\frac{4}{9}\theta^{2}\dot{G}f''+\frac{1}{3}\dot{G}^{2}f''^{2}-\frac{10}{9}\theta^{2}\dot{G}f''\;\nonumber\\
 && \quad\quad \nu=2H\dot{G}f''+\frac{G\dot{G}^{2}(f'')^{2}}{3H}\;.
\end{eqnarray}
We introduce dimensionless variables as:
\begin{eqnarray}
 && \mathcal{X}=\frac{Gf'-f}{6H^{2}}\;, \nonumber \\
 && \quad \quad \Omega_{m}=\frac{\rho_{m}}{3H^{2}}\;, \nonumber \\
 && \quad \quad \mathcal{Y}=6H\dot{G}f''.
\end{eqnarray}
We can present  different useful equations in redshift space as
\begin{eqnarray}
 &&\dot{H}=-\frac{2m}{3}(1+z)^{\frac{3}{m}}\nonumber\\
 &&\dot{G}=\frac{256}{9}m^{3}\Big(1-\frac{2m}{3}\Big)(1+z)^{\frac{15}{2m}}\nonumber \\
 && \ddot{G}=1280m^{3}\Big(\frac{2m}{3}-1\Big)(1+z)^{\frac{9}{m}}\nonumber \\
 && \dddot{G}=\frac{2560}{3}m^{3}\Big(1-\frac{2m}{3}\Big)(1+z)^{\frac{21}{2m}}.
  \end{eqnarray}

%\section*{Appendix}
 \noindent
{\color{blue} \rule{\linewidth}{1mm} }
  \end{document}